\documentclass[preprint2]{aastex}

\newcommand{\kms}{km s$^{-1}\;$}
\newcommand{\kmss}{km s$^{-1}$}

\newcommand{\lsun}{\mbox{L$_{\sun}$}}

\newcommand{\ho}{H$_{2}$O$\;$}

\newcommand{\mb}{mJy beam$^{-1}$}

\newcommand{\bm}{beam$^{-1}$}

\shorttitle{VLBI Observations of NGC~6240}
\shortauthors{Hagiwara, Baan, and Kl\"{o}ckner}

\begin{document}

%% LaTeX will automatically break titles if they run longer than
%% one line. However, you may use \\ to force a line break if
%% you desire.

\title{VLBI Observations of NGC6240: \\
    resolving the double nuclei and radio supernovae}

%% Use \author, \affil, and the \and command to format
%% author and affiliation information.
%% Note that \email has replaced the old \authoremail command
%% from AASTeX v4.0. You can use \email to mark an email address
%% anywhere in the paper, not just in the front matter.
%% As in the title, use \\ to force line breaks.

\author{Yoshiaki Hagiwara\altaffilmark{1}}
\affil{National Astronomical Observatory of Japan, 2-21-1, Osawa, Mitaka, 
181-8588 Tokyo, Japan}
\email{yoshiaki.hagiwara@nao.ac.jp}

\author{Willem A. Baan\altaffilmark{2}}
\affil{ASTRON, PO Box 2, 7990 AA Dwingeloo, The Netherlands}

\and

\author{Hans-Rainer Kl\"{o}ckner\altaffilmark{3}}
\affil{University of Oxford, Denys Wilkinson Building, Keble Road, Oxford OX1 3RH, 
United Kingdom}

\altaffiltext{1}{Department of Astronomical Science, The Graduate University for 
Advanced Studies (Sokendai), 2-21-1, Osawa Mitaka, 181-0015 Tokyo, Japan}
\altaffiltext{2}{Department of Mathematics, Physics and Computer Science, Linnaeus 
University, 351 95 V\"axj\"o, Sweden} 

\altaffiltext{3}{Max-Planck-Institut f\"{u}r Radioastronomie,
Auf dem H\"{u}gel 69, 53121 Bonn, Germany}

\begin{abstract}
The European VLBI Network (EVN) has been used at two epochs in 2003 and 2009 
to obtain multi-frequency high-resolution images of the merger galaxy NGC\,6240 in 
order to study the radio properties of all compact high-brightness components in the galaxy. 
Our observations at milli-arcsecond resolution {detected}
the northern and southern nuclei and two radio components, which we interpret as
long-lived luminous supernovae associated with the circum-nuclear starburst activity at the southern nucleus.
The new VLBI data support the presence of an active galactic nucleus (AGN) together 
with starburst activity at the southern nucleus and provides some evidence for an 
AGN at the northern nucleus. The two nuclei both display an inverted spectrum at lower GHz frequencies.
The spectrum of the southern nucleus indicates thermal free-free absorption on parsec scales, consistent with the presence of an AGN.

\end{abstract}

%% Keywords should appear after the \end{abstract} command. The uncommented
%% example has been keyed in ApJ style. See the instructions to authors
%% for the journal to which you are submitting your paper to determine
%% what keyword punctuation is appropriate.


%% their objects with \objectname{} or \object{}. 
%%
%% Note that for sources with brackets in their names, e.g. [WEG2004] 14h-090,
%% the brackets must be escaped with backslashes when used in the first
%% square-bracket argument, for instance, \object[\[WEG2004\] 14h-090]{90}).
%%  Otherwise, LaTeX will issue an error. 

%%%%%%%%%%%%%%%%%%%%%%%
\section{Introduction}
%%%%%%%%%%%%%%%%%%%%%%%

{The merging galaxy \object{NGC 6240} has a large
far-infrared (FIR) luminosity of L$_{\rm FIR}$ = 3.5 $\times$
10$^{11}$ \lsun\,\citep{yun02} and belongs to the category of luminous
infrared galaxies (LIRGs;  L$_{\rm IR}$ = 10$^{11}$$-$10$^{12}$ \lsun)
\citep[see][for a review]{sand96}}.
{The large luminosity of these LIRGs and Ultra-Luminous Infrared Galaxies 
(ULIRGs; L$_{\rm IR}$ = 10$^{12}$$-$10$^{13}$ \lsun)} is likely to be dominated by starburst
activity induced by galaxy-galaxy interactions, where the UV radiation from
massive star formation is reprocessed to far-infrared radiation in the
dusty environment \citep[e.g.,][]{sand88, sand96, ski97}.
The \object{NGC 6240} merging system has been described at optical 
wavelengths \citep{fri83} and is thought to host an active galactic nucleus (AGN) in its 
nuclear region \citep[e.g.,][]{dep86}.  The two galactic nuclei found in the central
region have a projected separation between
1$\arcsec$.5 and 1$\arcsec$.8 (0.714$-$0.856 kpc) at X-ray and radio wavelengths
\citep{fri83,bes01,kom03,gall04,max07}.  Because the northern nucleus, with the highest 
absorbing column densities, lies well behind the southern nucleus \citep{baa07}, the
separation varies as the intervening hot (obscuring) dust 
becomes optically thinner at longer wavelengths \citep{max07}. 
X-ray observations at 2$-$10 keV reveal the presence of iron line emission 
at 6.4 keV at both nuclei, which is most prominent in the southern nucleus 
and classifies this as a binary AGN  \citep{kom03}.

%Molecular lines
Thermal CO emission reveals a significant mass 
concentration that is centered between the two nuclei \citep{tac99,bry99,nak05,dai07} 
and largely consists of interstellar gas \citep{sco00}.
This intervening molecular gas concentration results from a superposition of disk gas from the 
two merging galaxies, which is confirmed by the structure and peak location of the OH absorption 
structure against the extended radio emission of the nuclear region \citep{baa07}. 

%VLA MERLIN
Radio observations at centimeter wavelengths of the central 2$\arcsec$ region of 
\object{NGC 6240} using the NRAO Very Large Array display the two 
nuclei and an extended northern structure (N3) that is part of the large-scale  
loop structure to the west of the nuclear source \citep{col94,baa07}.  
A large scale diffuse component envelops these three structural components. 
The nuclear radio components remain unresolved at higher resolution
at 5 GHz using MERLIN \citep{bes01}. The loop structures including N3
have been resolved away at this resolution and the southern nucleus, N1 shows east-west structure
at its lowest contours (PA = 60$\degr$) and a northern extension (PA = -30$\degr$).  

%VLBA
At milli-arcsecond  resolution using the NRAO Very Long Baseline Array (VLBA), 
the northern nucleus N2 also displays a multi-component east-west structure suggesting 
a core with a two-sided jet \citep{gall04}. 
The southern nucleus  has a peak central structure surrounded by weak 
additional components possibly related to starformation. Both high-brightness 
nuclei show an inverted spectrum at lower frequency 
with a spectral turn-over or flatness due to free-free absorption (FFA) or possibly
synchrotron-self absorption (SSA) \citep{gall04}. {Strong \ho maser emission was discovered 
\citep[e.g.,][]{hagi02} at the exact center of the southern nucleus \citep{hagi03,hagi10}, 
which may be related to a dense molecular gas around a central engine.}

Here, we present European VLBI Network (EVN) observations with increased 
surface brightness sensitivity to image the compact sources at increasing
frequencies. New EVN observations at 1.6, 2.2, 5.0 and 8.4 GHz were made to
study the source structure, the overall spectral energy distribution, and to 
check for consistency with the earlier VLBA results.
In Section 2 we present the new observations, the data analysis, and the observational 
results, {including the identification and morphology of all detected compact radio 
sources.} In Section 3 we discuss the 
nature of these sources and compare \object{NGC 6240} with other 
active galaxies. Finally, in Section 4 we summarize the discussion.

Cosmological parameters of H$_{0}$ = 73
\kmss Mpc$^{-1}$, $\Omega$$_{\Lambda}$ = 0.73, and $\Omega$$_{M}$ =
0.27 are adopted throughout this article, which results in a luminosity distance of
NGC 6240 of 103 Mpc at z = 0.0245 and an angular conversion of 1 arcsec equaling
476 pc in linear scale.  The designation of radio components follows that 
defined in \cite{col94} with N1 being the southwestern nucleus and N2 being the northeastern nuclei.
The two nuclei were called respectively S and N1 in \cite{gall04}.

%%%%%%%%%%%%%%%%%%%%%%
%\section{Observation}
%%%%%%%%%%%%%%%%%%%%%%%
\section{Imaging the double nuclei in NGC 6240}

\subsection{Observations and imaging}
%%%%%%%%%%%%%%%%%%%%%%%
%
The European Very Long Baseline Interferometry Network (EVN) has been used to
observe the nuclear region of \object{NGC 6240} in order to investigate the nature
of the active galactic nuclei and nuclear starbursts in the galaxy.  
The stations used for these observations at two epochs in 2003 and
2009 were Effelsberg (Ef), Hartebeesthoek (Hh), the 76-m Lovell
telescope (Jb) at Jodrell Bank, Medicina (Mc), Noto (Nt), Onsala
(On), Shanghai (Sh), Torun (Tr), Urumqi (Ur), and the Westerbork (Wb) 
phased-array.  A summary of the observational details is presented in Table~\ref{tbl1}.  
All observations were conducted in phase-referencing mode at all frequency bands
using the nearest calibration source J1651+0129  (centered at R.A.(J2000) 
= 16$^{\rm h}$51$^{\rm m}$03$^{\rm s}$.6620, Dec.(J2000) =
 01$\degr$29$\arcmin$23$\arcsec$.458) at
1.01$\degr$ away from \object{NGC 6240}.  Only the Lovell telescope did not
participate in the phase-referencing part of the observations due to its
lower slewing rate.  

The first epoch observations at 1.6 GHz and 5.0 GHz were made
on 30 October and 10 November, 2003 and utilized a 32 MHz total bandwidth with 2 bit
sampling in dual-circular polarization mode for a total data recording rate
of 256 Mbit s$^{-1}$.  Observations of 2.5$-$3 minutes on the target were
interleaved with observations of 1.5$-$2 minutes on the
phase-referencing source, which gave about 4.2 hours total on-source time.

Second epoch observations at 5.0 GHz, 2.2 GHz, and 8.4 GHz were 
made on 16, 17 June, 2009.  The observations at 5.0 GHz were made in 
dual-circular polarization mode utilizing 64 MHz total bandwidth with 2 bit 
sampling for a total data recording rate of 512 Mbit s$^{-1}$.  Three 
minute scans on the target were alternated with one minute scans on 
the phase-referencing source, which resulted in 5.7 hours total on-source 
time. The observations at 2.2 and 8.4 GHz were made in single polarization 
mode with the S/X dual-frequency receiving system with a bandwidth of 
128 MHz at each frequency and 2 bit sampling for a total recording rate 
of 512 Mbit s$^{-1}$.  Three minute scans on the target and one minute 
scans on the phase-referencing source were alternated during the 
observation, which resulted in 2.9 hrs total on-source time at each frequency.
All observations were done with the same reference position near the southern
nucleus of NGC 6240 (R.A.(J2000) = 16$^{\rm h}$52$^{\rm m}$58$^{\rm s}$.8903, 
Dec.(J2000) = 02$\degr$24$\arcmin$03$\arcsec$.339).  
 
The data were correlated using the EVN MkIV correlator at the Joint 
Institute for VLBI in Europe (JIVE) and the output visibility data were 
{averaged to} 1-2 seconds. Sixteen frequency channels of 500 kHz 
were used across each 8 MHz band. 
Data calibration and imaging were carried out in the NRAO Astronomical Image 
Processing System (AIPS). After initial data editing, the delay error corrections 
resulting from ionospheric effects were applied to all visibility data.  
A-priori amplitude calibration was performed using the system 
temperature and gain information provided for each telescope.  The delays 
and fringe rates for the calibration sources \object{3C 273} or \object{3C 345}, and 
\object{J1651+0129} were applied to the target source, \object{NGC 6240}. 
Using the measured flux densities of the phase-referencing source J1651+0129 at 2 epochs
as a basis, the relative accuracy of the target flux densities is estimated to be $\sim$10\%.
This error estimate includes the effects of bandwidth and time smearing and ionospheric delays, 
which are estimated to be several percent for these EVN observations.

Cleaned images of 512 $\times$ 512 pixels were produced, one centered 
near the southern nucleus and one at the intensity peak
position of the northern nucleus offset by 0$\arcsec$.5051 in
right ascension and 1$\arcsec$.4279 in declination from the phase
center.  The noise levels of 1$-$2 times theoretical noise and
synthesized beams from CLEAN-ed images are summarized in
Table~\ref{tbl2}. Self-calibration could not be applied successfully because of the low
signal-to-noise ratio of the images. Final images have been 
phase-referenced to the nearby calibrator \object{J1651+0129}. 
{Because of residual errors in the calibration of the phase-reference 
sources, the associated structure at the lowest contours ($\lesssim$ 3 $\sigma$ or 4 $\sigma$) in the source images may not be reliable.}

The positions of the radio components and their errors at the peak brightness 
at each frequency and both epochs are listed inTable~\ref{tbl3}. 
These composite position errors have been estimated from 
{the theoretical thermal noise errors of interferometer phase} \citep[e.g.,][]{tho86}, 
the systematic errors of the
phase-referencing VLBI observations ($\sigma_{\rm phr}$), and
the errors of the absolute position of the phase-referencing source ($\sigma_{\rm ref}$), using the quadrature equation:
$\sqrt{(\theta/2SNR)^2 + \sigma_{\rm phr}^2 + \sigma_{\rm ref}^2}$.
Here $\theta$ represents the synthesized beam size and SNR represents
the signal-to-noise ratio.  The values of $\sigma_{\rm phr}$ ($\Delta
\alpha \, $cos$\delta$\,=\,0.060 mas, $\Delta$$\delta$\,=\,0.020 mas) are
obtained from the simulated astrometric accuracy of the EVN array
\citep{pra06}, adopting a separation between calibrator and
target of 1.0$\degr$ at position angle 152.2$\degr$ and at an approximate
source declination of 0$\degr$. The value of $\sigma_{\rm ref}$ is 0.58 (mas) from the 
VLBA Calibrator List \citep{kova07}. Table~\ref{tbl4} lists the flux
densities of the radio components at each observing frequency,
which also includes two weaker components appearing in the earlier VLBA observations 
\citep[see Figures 2, 3, and 4 in][]{gall04}.  Source sizes of these components in 
Table~\ref{tbl5} are measured from the final images using the AIPS task JMFIT.
Assuming Gaussian distributions for sources, the upper limits of the source sizes have estimated 
accuracies of 3$-$30\% depending on observing frequency and epoch.
Table~\ref{tbl6} presents the properties of the
radio sources detected in our observations.
%
%%%%%%%%%%%%%%%%%%%%%%%%%%%%%%%%%%%%%%%%%%%%%%%%%
\subsection{Source identification and morphology}
\begin{figure}
\epsscale{0.8} \plotone{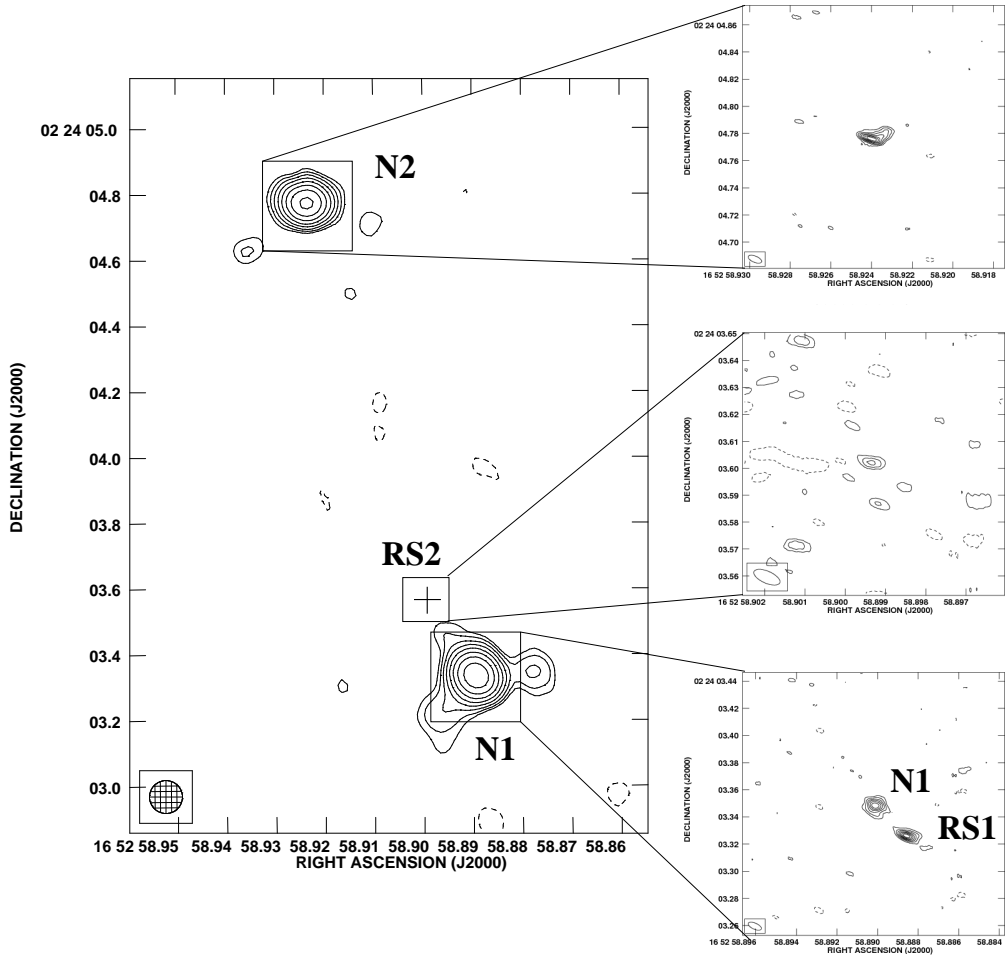}
\caption{The radio structure of the nuclear region of NGC 6240 at 5 GHz as obtained by
MERLIN \citep{gall04} with the location of compact components detected by the new 5 GHz EVN 
observations in October 2003 (Epoch 1) as presented in the right-hand-side frames.
The synthesized beam (FWHM) of the observations is plotted in the left bottom corner of all maps.
The two radio nuclei N1 (southwest) and N2 (northeast) have been labelled following the 
designation of Colbert et al. (1994) and Baan et al. (2007).  
{Note that in \cite{gall04} the northern nucleus is labeled as N1 and the southern nucleus 
labeled as S.}  The cross indicates the positions of radio source RS2
that appeared in our observation.
The contours in the MERLIN image are at 0.5 \mb $\times$ -1, 1, 2, 4,..., 32.
The contours of the three VLBI components are given with Figures 2, 3, and 4.
}
\label{fig1}
\end{figure}

\begin{figure}[htb]
\begin{center}
\epsscale{0.8}
\plotone{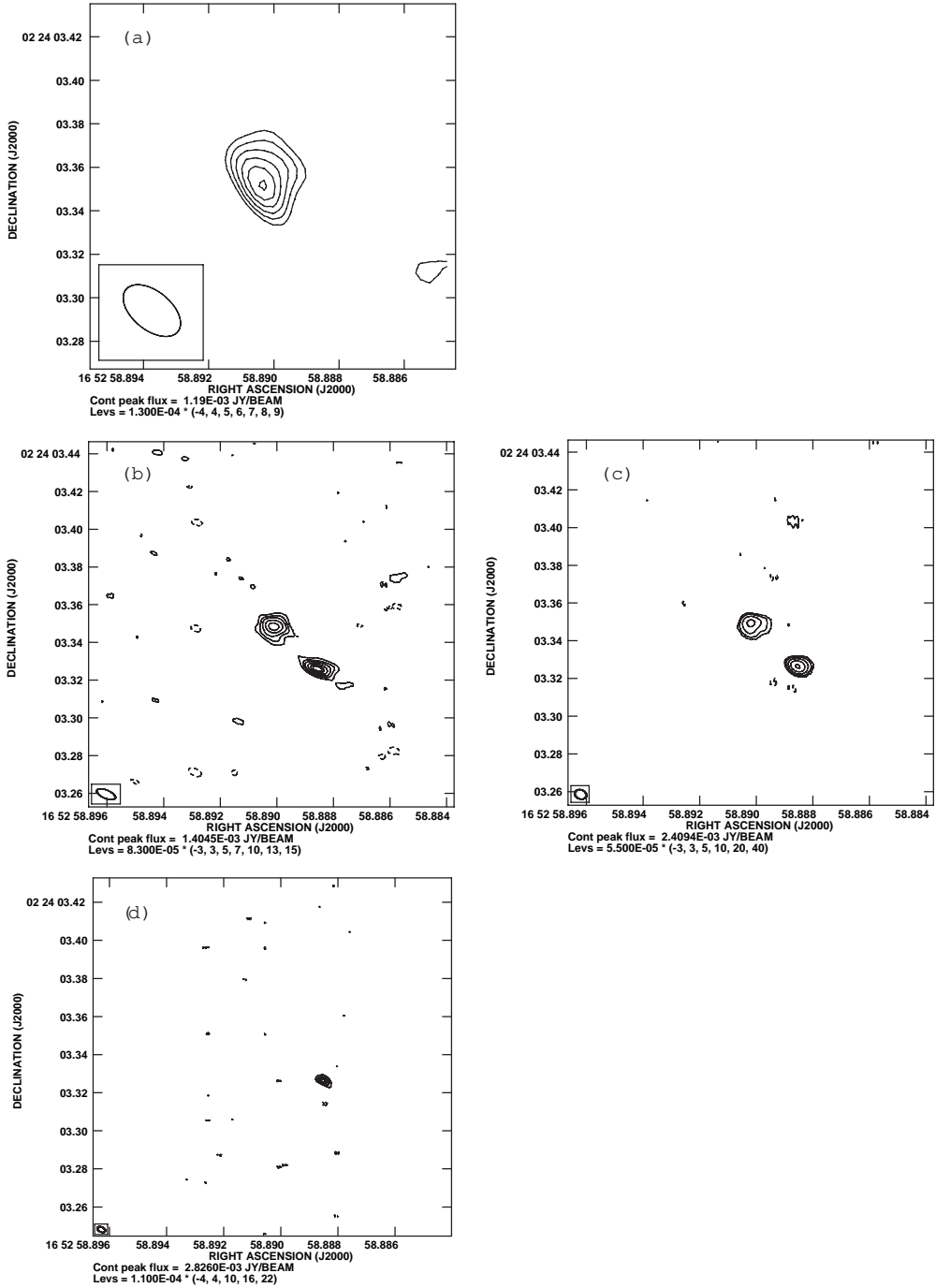}
\end{center}
\caption{Radio continuum images of the southwestern nucleus N1 of NGC\,6240 and the 
compact radio source RS1, which is located southwest of N1. 
The synthesized beams are plotted at the bottom-left corner on each panel.
(a) The 1.6 GHz image of N1 observed in Epoch 1 in 2003.
(b) The 5.0 GHz images of N1 and RS1 observed in Epoch 1 in 2003.
(c) The 5.0 GHz images of N1 and RS1 observed in Epoch 2 in 2009.  
(d) The 8.4 GHz image of RS1 obtained in Epoch 2.
%Contours are -4,4,5,7,10,15 $\times$1$\sigma$
}
\label{fig2}
\end{figure}

%%%%%%%%%%%%%%%%%%%%%%%%%%%%%%%%%%%%%%%%%%%%%%%%%%
%
The new VLBI maps obtained from 1.6$-$8.4 GHz display four compact
high-brightness emission components as identified in Figure \ref{fig1}: the 
nuclei, N1(south) and N2(north) identified in earlier interferometric 
observations \citep{bes01,hagi03} and two new components RS1 and RS2
near the southern nucleus. The radio source RS1 was first identified southwest of 
the southern nucleus at 5.0 GHz in October, 2003 and it is detected at 5.0 
and 8.4 GHz in June, 2009 (see Figure~\ref{fig2}). The other radio source 
RS2 was identified northeast of the southern nucleus at Epoch 1 and was 
more prominent at 1.6 GHz than at 5.0 GHz (see Figure~\ref{fig4}). 
At Epoch 2 {the source RS2 was clearly detected} at 5.0 
GHz but not at 2.2 GHz with an upper limit flux density of 2.75 \mb.
All diffuse emission in the central 2$\arcsec$ region, {observed in earlier 
Very Large Array observations \citep[e.g,][]{col94,baa07}}, has been resolved out.  

\begin{figure}[htb]
\begin{center}
\epsscale{0.8}
\plotone{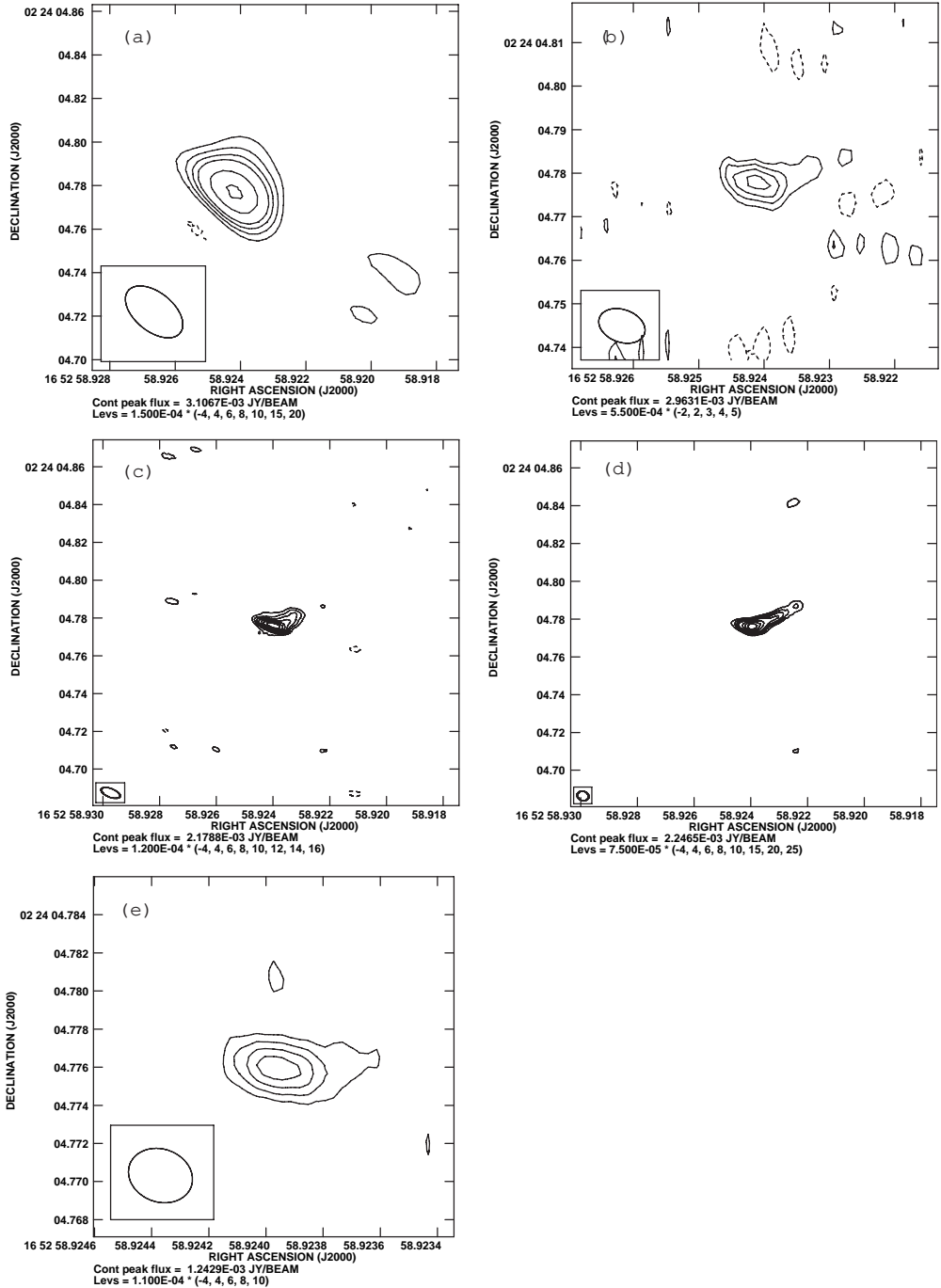}
%\plottwo{f3a.eps}{f3b.eps}
\end{center}

\caption{Radio continuum images of the northeastern nucleus N2 of NGC\,6240.  
The synthesized beams are plotted at the bottom-left corner on each panel. 
(a) The 1.6 GHz image observed in Epoch 1 in 2003.
(b) The 2.2 GHz image observed in Epoch 2 in 2009.
(c) The 5.0 GHz image observed in Epoch 1 in 2003.  
(d) The 5.0 GHz image observed in Epoch 2 in 2009. 
(e) The 8.4 GHz image  obtained in Epoch 2 in 2009.
}
\label{fig3}
\end{figure}

\begin{figure}[htb]
\begin{center}
\epsscale{0.9}
\plotone{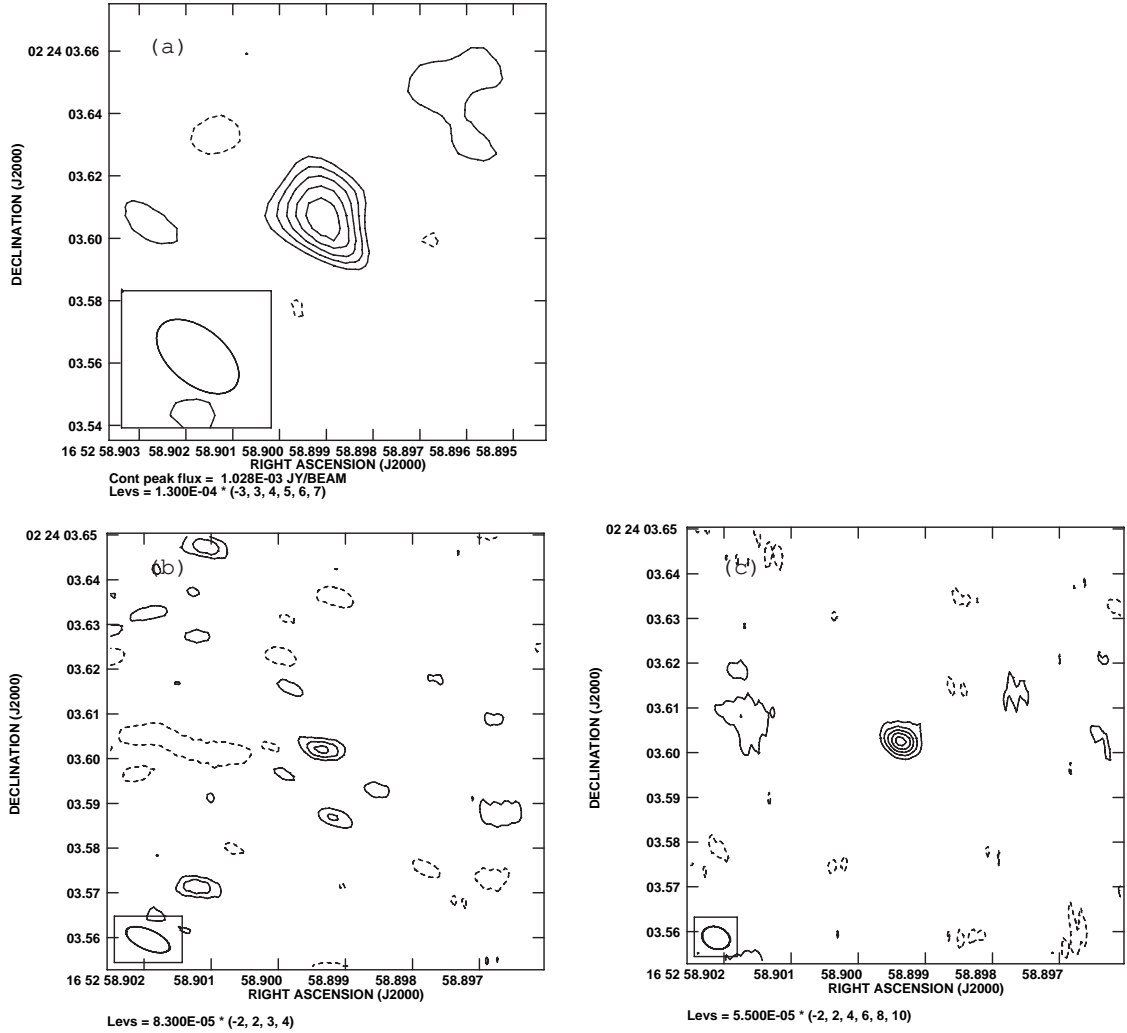}
\end{center}
\caption{ Radio continuum images of the radio source RS2 that {is located} northeast 
of the southern nucleus N1 as indicated in the lower angular resolution map obtained by 
the 5 GHz MERLIN (Figure~\ref{fig1}). The synthesized beams are plotted at the 
bottom-left corner on each panel.  (a) The 1.6 GHz image obtained in Epoch 1 in 2003.
(b) The 5.0 GHz image of Epoch 1 in 2003. (c) The 5.0 GHz image of Epoch 2 in 2009.}
\label{fig4}
\end{figure}

The VLBA observation at 8.4 GHz of August 1999 did not detect 
the two nuclei at a 5$\sigma$ upper limit of 0.65 \mb \citep{gall04}, but 
the northern nucleus was detected at 8.4 GHz in June 2009 with similar
sensitivity, which suggests some intensity variability of the northern nucleus. 
Both nuclei were detected at 5 GHz at both Epochs. Their peak positions in
 the EVN data (Table~\ref{tbl3}) agree at all frequencies with 
earlier VLBA measurements using estimated uncertainties of 0.6 mas in 
right ascension and 1.5 mas in declination  \citep{gall04}.  
The 5.0 GHz maps in Figures~\ref{fig2} and \ref{fig3} display the two nuclear 
components N1 and N2 obtained at Epoch 1 and 
Epoch 2, respectively. The southeastern nucleus, N1 was not detected at the highest 
frequency of 8.4 GHz with an upper limit of 0.55 \mb (Figure~\ref{fig2}). Individual source 
maps indicate that {the sources N1 and N2 are resolved but that the minor axis of 
RS2 northeast of N1 is unresolved (Figure~\ref{fig4})}. The peak of RS1 is 
separated from N1 by 23.3 mas at 5.0 GHz at a position angle of $\sim$35$\degr$, 
corresponding to 11.1 pc at the distance to NGC 6240. 
At the lowest frequency of 1.6 GHz, N1 and RS1
are not spatially separated.  Also, RS1 was not detected at 2.2 GHz
at Epoch 2 possibly because of the lower signal-to-noise ratio. 

The source RS2 is located northeast of the southern nucleus in our 1.6 GHz 
EVN map of Figure~\ref{fig4}. It appears that the
peak position of RS2 at 1.6 and 5.0 GHz lies close to the compact component S1 at
0$\arcsec$.3 northeast from N1 in the 1.7 and 2.4 GHz VLBA observations 
\citep{gall04}. On the other hand, RS2 has an offset of $\approx$ 10$-$15~mas 
at 1.6 GHz or 5.0 GHz relative to S1 at 1.7 GHz or 2.4 GHz, which is 
not within the estimated positional errors (Table~\ref{tbl3}).
Therefore, RS2 is a different source to S1 in the earlier VLBA data.
We interpret both components to be part of the 
circum-nuclear starforming region that extends over 10 mas or $\sim$5 pc around N1.

%N2
The structure of the northeastern nucleus N2 at 5.0 GHz  (Figure~\ref{fig3}) 
shows a northwest extension of $\approx$ 0\arcsec.02 ($\approx$ 9.5 pc) that 
is consistent with the earlier VLBA maps at 1.7 and 2.4 GHz \citep{gall04}.  
Although {this structure at lower contours} may not be reliable because of 
incomplete phase and amplitude calibration, an east-west structure may be 
explained as a core-jet structure typically seen in high-luminosity Seyfert nuclei.  
The northern nucleus itself remains unresolved at both 1.6 and 2.2 GHz.
%8.4GHz
Only at the highest frequency of 8.4 GHz, N2 has an east-west elongation that 
is consistent with the earlier VLBA data (Figure~\ref{fig3}).  

%
%
%%%%%%%%%%%%%%%%%%%%%%
\section{Discussion}
%%%%%%%%%%%%%%%%%%%%%
%
%\subsection{Properties of the double nuclei}
%
Our EVN data present the detection of the double nuclei at higher 
frequencies of 5.0 and 8.4 GHz, where earlier VLBA data failed to 
detect these nuclei. These results combined with earlier VLBA studies,
allow study of the nature of the nuclei at milliarcsecond-scale structure.  
While our study focuses on the physics of the two nuclei, the detection 
of two new components is a key to understanding the activity of NGC 6240.
%
%%%%%%%%%%%%%%%%%%%%%%%%%%%%%%%%%%%
\subsection{The northern nucleus N2}
%%%%%%%%%%%%%%%%%%%%%%%%%%%%%%%%%%%
%
The northern nucleus N2 shows an inverted spectrum rather than a steep spectrum 
that would be more typical for a radio jet (Figure~\ref{fig5}). The structure of N2 at 5.0 GHz is 
partially resolved and shows an east-west extension at its lower 
contours that still remains questionable because of the insufficient amplitude 
calibration (Figure~\ref{fig3}). Changes in source structure may also result
from a different (u,v) coverage for the two epochs. The detection of the X-ray emission 
towards N2 \citep{kom03} is less convincing than the definite detection towards N1. 
Nevertheless, the radio morphology of N2 resembles that of a core-jet 
structure seen in Seyfert nuclei, such as \object{NGC 1068} \citep{gall04b}, but 
this requires further verification. 
By contrast with \object{Arp 299}, the VLA observations of the \ho maser in
NGC 6240 have not detected any emission towards N2 during seven years
\citep{hagi03,hagi10}.
%
%%%%%%%%%%%%%%%%%%%%%%%%%%%%%%%%%%%%%
\subsection{The southern nucleus N1}
%%%%%%%%%%%%%%%%%%%%%%%%%%%%%%%%%%%%
%
The southern nucleus N1 appears partially resolved in an east-west
direction (Figure~\ref{fig2}b,2c) but its flux density at 5.0 GHz has
not changed significantly during 6 years.  Contrary to the spectrum of N2
(see discussion below), the spectrum of N1
with missing Epoch 2 detections at 2.2 and 8.4 GHz may not be inverted
(Figure~\ref{fig5}), although its spectral shape (including
upper limits) strongly indicates a spectral turnover around 2.0 GHz.
The non-detection of N1 at 8.4 GHz suggests a relatively flat AGN-like 
spectrum and {a lower limit for the spectral index $\alpha$ $<$ 0.55  
between 5.0 and 8.4 GHz, using S$_{\nu}$ $\propto$ $\nu^{\alpha}$ with 
S$_{\nu}$ being the flux density at frequency $\nu$}. In addition, the radio brightness 
temperature $>$ 10$^6$ K (Table~\ref{tbl6}) and the detection of strong hard 
X-ray emission and neutral iron line emission would argue for the presence 
of an AGN in N1.  Considering the available evidence, the southern
nucleus may host both an AGN and a (circum-)nuclear starburst region. 

The \ho maser features in NGC 6240 nearly coincide with the 
continuum peak of the southern nucleus \citep{hagi03,hagi10}. 
The narrow \ho maser lines are redshifted by $\sim$ 200$-$300 \kms 
with respect to the systemic velocity of N1 \citep[e.g.,][]{hagi10,baa07} 
and may  originate in the receding side of a compact rotating molecular disk 
at the nucleus \citep[e.g.,][]{miyo95}, which would also support the presence 
of an AGN. Alternatively, these narrow lines distributed over 120 \kms could 
be explained by wind maser emission as observed in the Circinus galaxy 
\citep{linc03}.

%The \ho maser features in NGC 6240 have been associated with the 
%continuum peak for the southern nucleus \citep{hagi03,hagi10}. However, the 
%blueshifted velocity of the two features may suggest that these are not nuclear maser 
%features associated with an AGN but rather with amplification related to the 
%circum-nuclear starformation. 

The southern nucleus N1 of NGC 6240 has characteristics that are comparable 
with those of component A of Arp\,299 \citep{nef04,ulv09}. 
The 8.4 GHz radio power of N1 of ($P_{N1}$ =  3.7 $\times$ 10$^{22}$ W Hz$^{-1}$)  \citep{col94}
is only a factor of two more than the 8.4 GHz
power of nucleus A of Arp\,299 ($P_A$=1.8 $\times$ 10$^{22}$ W 
Hz$^{-1}$ and $P_{B1}$=2.9 $\times$ 10$^{21}$ W Hz$^{-1}$) \citep{nef04}.
The compact source A of Arp\,299 has been resolved into discrete sources at 
milliarcsecond-scale by recent EVN observations and one of these sources (A1) 
is the AGN candidate with a flat spectral index of -0.13 $\pm$ 0.11 \citep{per10}.
Five bright and compact radio sources lie within 10 pc and are candidates of 
Type-II young radio supernovae  \citep{nef04}. Four of these five
radio sources have flat or inverted radio spectra between 2.2 and 8.4
GHz, very similar to RS1 and RS2 near N1 in NGC 6240.
The luminous \ho maser emission in Arp\,299 is found towards both nuclei A 
and B1 \citep{hen05} and at A it coincides with its continuum peak at VLA resolution 
\citep{tar10}. While this spatial coincidence may also suggest an AGN 
association \citep[e.g.,][]{hagi07}, the maser in \object{Arp 299} is blueshifted 
relative to the systemic velocity and is likely associated with an outflow in a starforming environment.

%%%%%%%%%%%%%%%%%%%%%%%%%%%%%%%%%%%
\subsection{Spectral analysis of N1 and N2}
%%%%%%%%%%%%%%%%%%%%%%%%%%%%%%%%%%%

In order to understand the observed spectra of the southern and northern 
nuclei, model fitting may be used to explain the frequency turn-over at lower 
frequencies using a power-law plus pure free-free absorption (FFA) or 
synchrotron self-absorption (SSA)  \citep[e.g.,][]{kam00}. First, 
the FFA model for a nuclear spectrum is described by:
\begin{equation} 
\label{ffa} 
S_\nu=S_{\rm 0}(\nu/1.0)^{\alpha_0} \exp\{{-\tau_{\rm ff}(\nu/1.0)^{-2.1}\}},
\end{equation} 
where $\nu$ is the frequency in GHz, $S_{\rm 0}$ is the unobscured
synchrotron flux density in mJy, $\alpha_0$ is the optically thin
non-thermal spectral index, {and $\tau_{\rm ff}$ is the opacity at 1.0 GHz. } 
This model describes all FFA in the foreground to the synchrotron
emission source and assumes that all source components are subject 
to the same foreground opacity \citep[e.g.,][]{par07}. The second model of 
pure SSA is described as follows:
\begin{equation} 
\label{ssa}
 S_\nu=S_{\rm 0}\nu^{2.5} [1-\exp\{{-\tau_{s}\nu^{\alpha_0-2.5}\}}],
\end{equation}
where $\nu$ is the frequency in GHz, $S_{\rm 0}$$\tau_{s}$ will be
close to the flux density, if the SSA coefficient $\tau_{s}$$<<$ 1, and
$\alpha_0$ is the optically thin non-thermal spectral index.

\begin{figure}[htb]
\begin{center}
\epsscale{1.1}
\plottwo{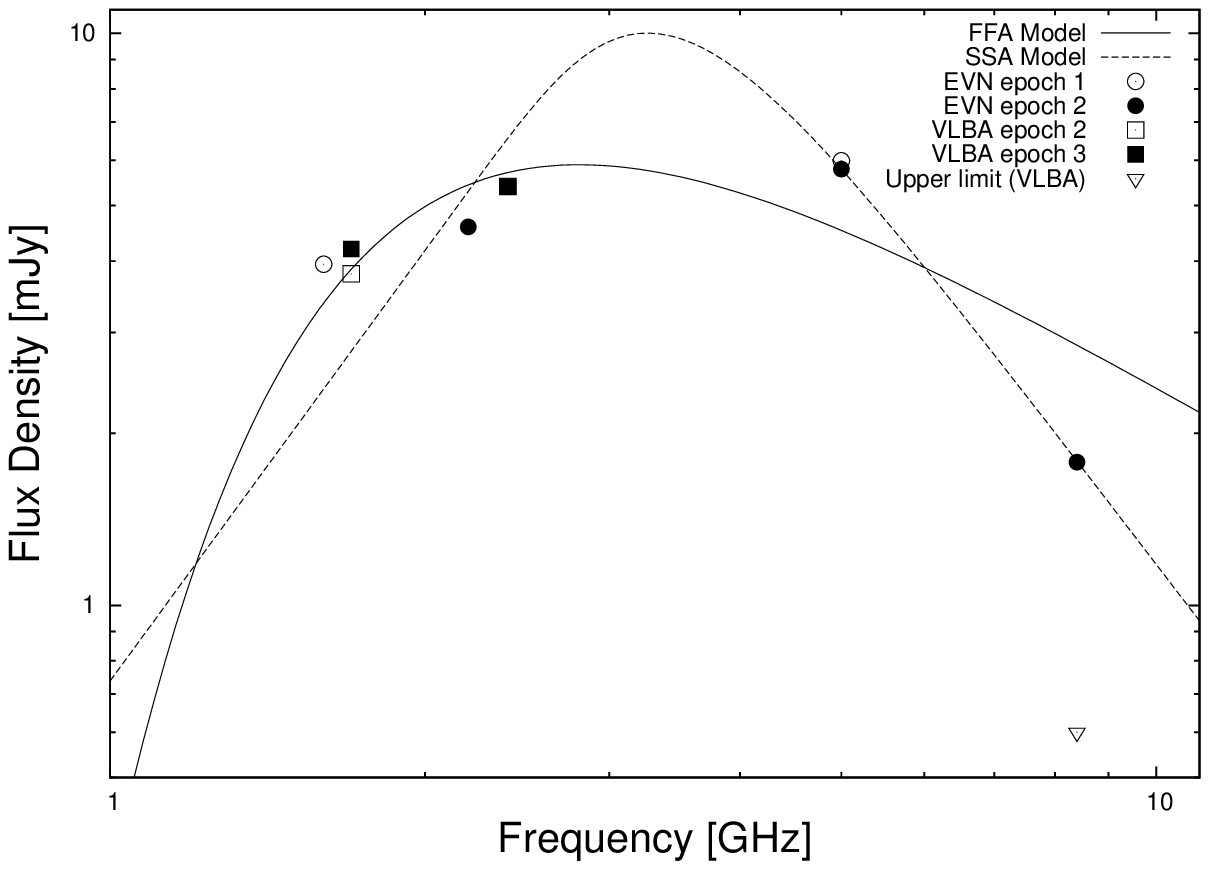}{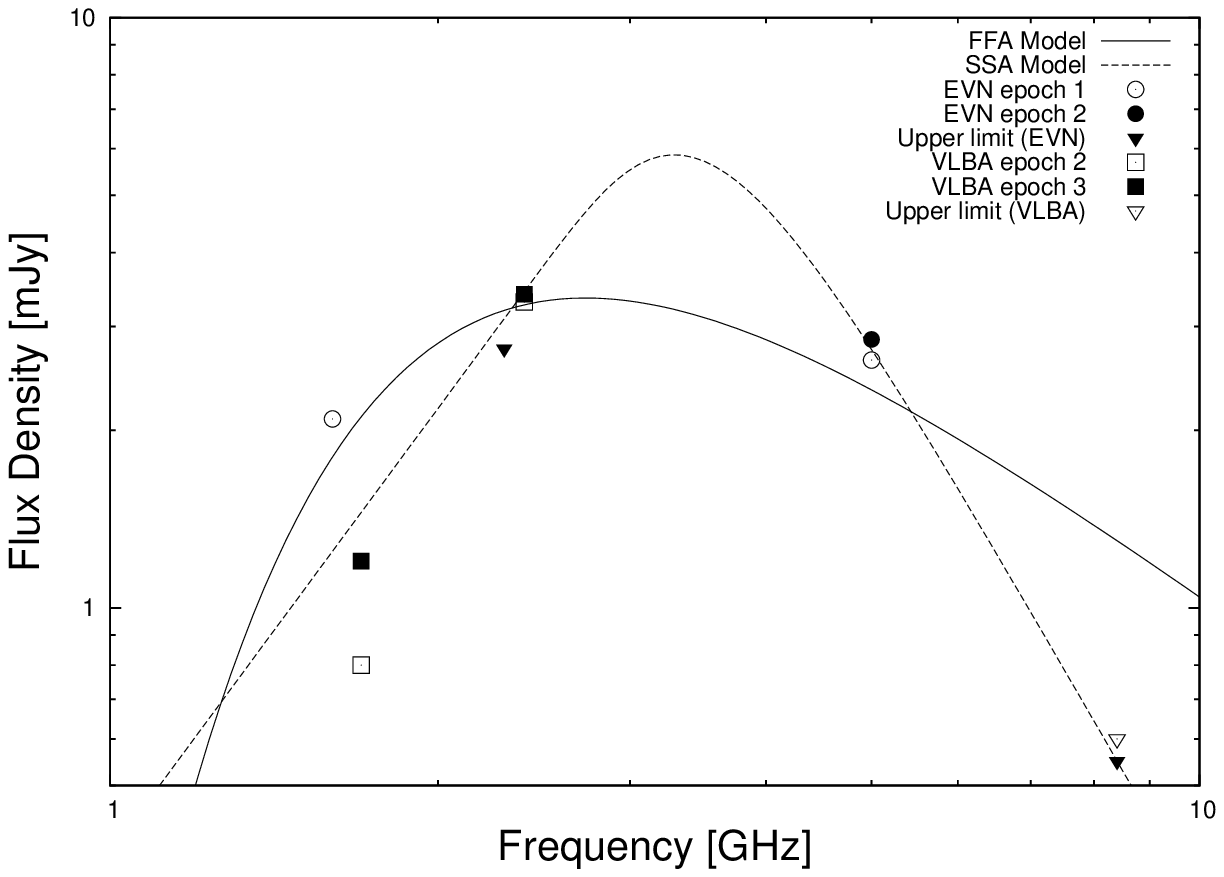}
\end{center}

\caption{Radio spectra of the two nuclei of NGC 6240.
    \newline {{\it Left frame}}: The spectrum of the northern nucleus N2. The solid lines represent 
    the free-free absorption (FFA) model as fitted to the Epoch 2 data and 1.6 GHz
    data in Epoch 1, assuming no significant flux density variation of N2 between 
    the two observing epochs.  The dotted line represents the synchrotron
    self-absorption (SSA) model as fitted to the same data as the case
    of FFA.  
    \newline {{\it Right frame}}: The spectrum of the southern nucleus N1.  The solid and dotted lines 
    represent the FFA model and SSA model as fitted to data of the two EVN epochs 
    including upper limit values, assuming no significant flux density variation of N1 
    during these two epochs.   
    \newline {\it The data points}: The filled circles at 2.2, 5.0, and 8.4 GHz are data points from the
    EVN during Epoch 2 (2009). The open circles at 1.6 and 5.0 GHz are data points
    from the EVN during Epoch 1 (2003).  The open squares at 1.7 and 2.4 GHz
    are the data points from the VLBA at Epoch 2 (April to May 2000),
    and filled squares at 1.6 and 2.4 GHz are from the VLBA data of
    Epoch 3 (December 2000 to January 2001).  The triangles
    denote the 5 $\sigma$ upper limit values obtained at 2.2 and 8.4 GHz
    with the VLBA in August 1999 from \cite{gall04}. 
    }
\label{fig5}
\end{figure}

The parameters of the FFA and SSA spectral fits for N1 and N2 and the 
reduced $\chi^2$ values (per degree of freedom) have been summarized 
in Table~\ref{tbl7}, and the fitted curves are shown in Figure~\ref{fig5}. 
Both the FFA and SSA models produce reliable fits for N1 with low values ($<$ 1.5)
for the reduced $\chi^2$ with the FFA model being slightly worse but
does not make good fits to N2. 
However, since there is no compact source in N1 or N2 having a high 
brightness temperature, any synchrotron
self-absorption in these nuclei is unlikely \citep{kel69}. Using conventional 
energy equipartition arguments at lower radio frequencies, a turn-over 
frequency of 2.0 $-$ 3.0 GHz for a self-absorbed source requires an equipartition brightness
temperature of $\sim$10$^{11.95}$ K \citep{rea94}. The observed 
brightness temperatures for our sources are only
$\sim$ 10$^{6}$$-$10$^{7}$ K. 
Better constraints of the nuclear properties of N1 in terms of model 
fitting requires more complete flux density measurements.

%On the other hand, the interpretation of the northern nucleus N2 is not 
%straightforward. 
While for N2 neither the pure FFA nor the pure SSA model can perfectly explain the spectral 
bending at 2$-$3 GHz (see Table~\ref{tbl7} and Figure~\ref{fig5}), the observed 
bending does indicate 
that free-free self-absorption by ionized foreground gas in a starburst environment 
is relevant. Since there is also no hint of self-absorption in the MERLIN 
spectrum \citep{bes01}, the size of such an absorbing medium must be very 
small and between about 9.5 and 25 pc in linear size  \citep{gall04}. 
Similarly, the presence of an AGN in N2 cannot be confirmed from our EVN data, 
except that the core-jet like structure, as also seen in the 1.7 and 2.4 GHz VLBA 
images, could support the presence of an active nucleus.

%and the RS2 has a relatively flat spectrum with 
%$\alpha_{\rm{1.6-5.0}}$ = -0.91, which suggests a similar origin to RS1.
%%%%%%%%%%%%%%%%%%%%%%%%%%%%%%%%%%%%%%%
\subsection{Radio sources near the southern nucleus RS1, RS2}
\begin{figure}[htb]
%\begin{center}
\epsscale{0.8}
\begin{center}
\plotone{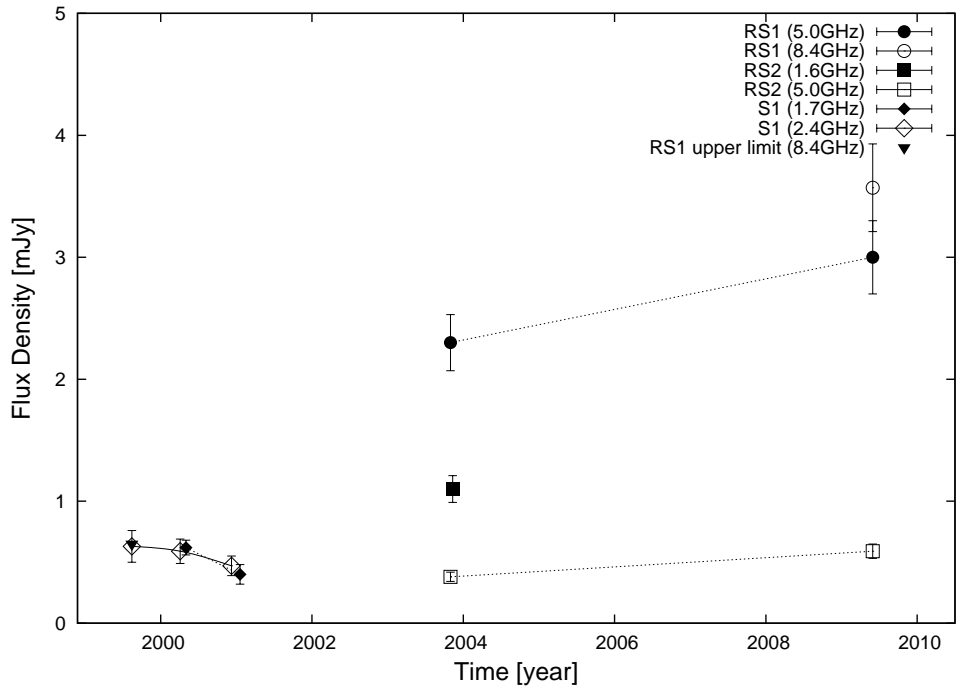}
\end{center}
\caption{Radio light curves of RS1, RS2, and S1. Filled circles denote the 5.0 GHz flux density of RS1, 
and an open circle denotes the 8.4 GHz flux density of RS1. A filled square shows the 1.6 GHz flux of RS2, 
and the open squares denote the 5.0 GHz flux of RS2. Filled and open diamonds denote the 1.7 GHz and 
2.4 GHz flux of S1 from the earlier VLBA data.
A open triangle indicates the upper limit (5 $\sigma$) of the 8.4 GHz earlier VLBA data.}
\label{fig6}
%\end{center}
\end{figure}

%%%%%%%%%%%%%%%%%%%%%%%%%%%%%%%%%%%%%%%
The radio source RS1 has a radio power at  8.4 GHz of 4.1 $\times$ 10$^{21}$
W Hz$^{-1}$ and a spectral index of $\alpha_{\rm{5.0-8.4}}$ $\approx$ 0.34 in 
our Epoch 2 observations. The radio power of RS2 at 5.0 GHz ($\alpha_{\rm{1.6-5.0}}$ $\approx$ -0.91) is 7.5 $\times$ 10$^{20}$ W Hz$^{-1}$. 
The radio power of RS2 is comparable with those of the VLBI sources associated 
with the nucleus A in \object{Arp 299} while the power of RS1 is nearly a factor 
ten higher, which makes it 1250 times the 8.4 GHz radio power of \object{Cas
A} of 6 $\times$ 10$^{17}$ W Hz$^{-1}$ \citep{col94}. 
Both RS1 and RS2 display a radio light curve with a long-term rise (Figure~\ref{fig6}). 
The {upper} limit for the source size of RS1 at the highest frequency 8.4 GHz is 1.5
$\times$ 0.6 pc (Table~\ref{tbl5}) and the source is not sufficiently
resolved by our EVN synthesized beam.  Also, different EVN beams resulting
from different (u,v) coverage at each epoch make it difficult to make a reliable 
comparison of the intrinsic source sizes.

Speculation that RS1 could be the true southern nucleus, instead 
of N1, may be ruled out because RS1 was not detected in the 8.4
GHz VLBA observations (1999$-$2001) and only appeared in 2003. Also its 
location is significantly offset from the positions of N1 in the 2.4 and 1.7 GHz VLBA data.   
Similarly, the possibility that RS1 represents ejecta from the active nucleus N1 
may be ruled out because its offset (11 pc) from N1 remained unchanged during 
6 years and because RS1 is far from the known outflow structures seen in HI and 
OH southwest of N1 \citep{baa07}. The positional difference of 10 mas ($\approx$ 
50 pc) (see Table~$\ref{tbl3}$) between RS2 at 1.6 and 5.0 GHz and the source S1 
at 1.7 and 2.3 GHz in VLBA observations at epochs in 2000 and 2001 
suggests that they are also different sources. 

The radio sources RS1 and RS2 have remained detectable for more than 6 years since 
2003. An important question is whether these sources are supernovae  (SNe, interacting with the circumstellar matter) or supernova 
remnants (SNRs, interacting with the dense ISM).
There is no evidence for expanding supernovae shells with the highest beam size of $\sim$6 mas ($\sim$3 pc) both 
in RS1 and RS2 between the two epochs from our data (see Table~$\ref{tbl5}$). The range of radio luminosity 
of 10$^{20.8-21.6}$ W Hz$^{-1}$ of RS1 and RS2 is similar to the radio luminosity obtained for observed Type Ib/c or Type II radio supernovae (RSNe) \citep{wei02}. Thus, it is possible that 
RS1 and RS2 are SNe or SN-SNR transition objects.
Our data is insufficient to distinguish between the two cases and further sensitive VLBI monitoring would be required.
The VLBA data (1999$-$2001) also displays two similar but weaker 
RSN or SNR candidates RS3 and RS4, the latter of which has a spectral index of
$\alpha_{\rm.{1.7-2.4}}$ $\approx$ 0.35. 

%Their relatively flat spectrum and long-lived activity would indicate 
%that they are young Type-II supernovae \citep{wei02} resulting  from the circum-nuclear 
%starburst at N1. 
Many nearby galaxies, such as the LIRG \object{M 82}, and ULIRGs \object{Arp 220} and 
\object{Arp 299}, display compact radio sources that are evidence of ongoing
(circum-)nuclear starformation \citep{smi98,mcd01,nef04,par07,con07,ulv09}.  
VLBI observations of the merging galaxy Arp\,220 have detected a total of 18 
compact radio sources within the western nucleus of the galaxy, over half of 
which have radio properties that are consistent with Type-II supernovae 
interacting with the surrounding medium \citep{par07}. Likewise, the nuclei in 
Arp\,299 show 30 compact radio sources, 25 of which are associated with the 
northeastern nucleus A (with an AGN candidate) and are spread over a region 
of 30 pc \citep{nef04,ulv09}. 

A comparison of the SN$-$SNR sources in NGC\,6240 with those found in other
sources shows that the radio powers of RS1 and RS2 are equivalent to those of
the most powerful sources found in M\,82 \citep[see][]{fenech08},
although they are less powerful than those in Arp\,220. As a result the upper limits for the sizes of RS1 and RS2 lie in the upper range of the general relation between radio luminosity and diameter observed for Galactic and extragalactic SNR sources \citep{ber04,bat10}. This would suggests that the environmental 
conditions in the LIRG NGC\, 6240, 
and possibly the star formation initial mass function, cannot yet be distinguished from those of the most luminous FIR galaxies.

%%%%%%%%%%%%%%%%%
\section{Summary}
%%%%%%%%%%%%%%%%%%%
We have conducted multi-frequency EVN observations of the nuclear
region of the merging galaxy NGC 6240 at two epochs.  The
new VLBI maps reveal the double radio nuclei of NGC
6240 at milliarcsecond resolution, that are consistent with the
earlier VLBI images obtained with the VLBA \citep{gall04}.  {The radio
spectra from both nuclei suggest a spectral turn-over between 2 and 3 GHz. The spectrum of
the southern nucleus N1 may be explained in terms of free-free absorption, 
although this explanation is still limited by having source flux densities with insufficient frequency coverage at each epoch. There 
is no clear interpretation for the spectrum of the northern nucleus N2. }

Questions still remain about the true nature of the two radio nuclei and whether they 
both contain a radio-quiet AGN, a simple starburst, or a composite with an  AGN and
a circum-nuclear starburst.  We suggest that the southern nucleus hosts
an AGN and a circum-nuclear starburst, as evidenced by the X-ray data and 
the radio sources. The association of the \ho maser with the nuclear source 
is still unknown.   However, it is not clear that our data, together with
earlier VLBA measurements, confirm the presence of an AGN at the northern
nucleus.

The radio components RS1 and RS2 could be interpreted as long-lived 
radio supernovae that result from ongoing circum-nuclear star formation at N1. The radio spectrum of both sources is relatively flat and their location 
remains unchanged within error
over about 6 years.  More radio supernovae may have been
detected in earlier observations around the active nucleus of N1,
which groups \object{NGC 6240} together with other well-known starburst
nuclei that display RSNe or SNRs in their nuclei, such as \object{M
  81},  \object{M82}, \object{Arp 220} and \object{Arp 299} \citep{bar09}. Radio interferometric
observations are a powerful method to identify ongoing nuclear
starformation and detect radio supernovae
and supernova remnants  in extragalactic sources.

\acknowledgments We are grateful to Drs. Yoshiharu Asaki and Seiji
Kameno for their helpful suggestions, and we also thank to Dr. Robert
Beswick for providing the MERLIN radio image.  The authors thank 
Dr. Bob Campbell and other staff members in JIVE for their
assistance in the observations, correlation, and data analysis.  
The authors also wish to thank an anonymous referee for suggestions that
improved the manuscript. YH acknowledges Dr. Phil Edwards and 
staff members in VLBI Space Observatory Programme 2 (VSOP-2) for 
providing valuable advice.
The European VLBI Network is a joint facility of European, Chinese, South
African and other radio astronomy institutes funded by their national
research councils.  This work was supported in part by The Graduate
University for Advanced Studies (Sokendai).  This research has made
use of NASA's Astrophysics Data System Abstract Service.
This research has made use of the NASA/IPAC Extragalactic Database (NED), 
which is operated by the Jet Propulsion Laboratory, California Institute of 
Technology, under contract with the National Aeronautics and Space 
Administration. 

\onecolumn
\clearpage
%
%
% ---------------------- TABLE 1-------------------------
\begin{table}
\small
%\begin{center}
\caption{EVN Observations of NGC 6240 \label{tbl1}}
\begin{tabular}{ccccccc}
\tableline\tableline
Epoch&Observing & Frequency &Telescopes & Data rate& Duration &On-source \\
        &Date            & (GHz)  &   & (Mbit s$^{-1}$) & (hrs) &(hrs) \\ 
%&\multicolumn{1}{c}{$P$\tablenotemark{a}} & $P R_{maj}$ & $P R_{min}$ &
%&\multicolumn{1}{c}{$\Theta$\tablenotemark{b}} \\
\tableline
1&30 Oct  2003&4.97&Ef,Hh,Mc,Nt,On,Wb&256&7.9& 4.2\\
1&10 Nov 2003&1.64&Ef,Hh,Jb,Mc,Nt,On,Wb&256&8.2& 3.6\\
\tableline
2&16 Jun 2009&4.96&Ef,Jb,Mc,Nt,On,Sh,Tr,Ur,Wb&512&8.9&5.7\\
2&17 Jun 2009&2.24&Ef,Mc,Nt,On,Sh,Ur,Wb&512\tablenotemark{a}&8.7&5.8\\
2&17 Jun 2009&8.36&Ef,Mc,Nt,On,Sh,Ur&512\tablenotemark{a}&8.7&5.8\\
\tableline
\end{tabular}
%% Any table notes must follow the \end{tabular} command.
\tablenotetext{a}{Total recording rate at both S and X band.}
%generated with the \LaTeX\ table environment}
%\tablenotetext{b}{Yet another sample footnote for table~\ref{tbl-2}}
%\tablenotetext{c}{Another sample footnote for table~\ref{tbl-2}}
%\tablecomments{We can also attach a long-ish paragraph of explanatory
%material to a table.}
%\end{center}
\end{table}

%------------------------ TABLE 2--------------
\begin{table}
\small
%\begin{center}
\caption{VLBI Imaging Parameters \label{tbl2}}
\begin{tabular}{ccccc}
\tableline\tableline
Epoch & Frequency  & Rms (Uniform-weight)     &Beam size (FWHM) & PA  \\
         &(GHz)          & (mJy \bm)     &   (mas)& ($\degr$) \\ 
\tableline
1 &1.6&0.13& 30.8  $\times$ 17.8   &50.9\\
1 &5.0& 0.083 &10.5 $\times$ 4.58&50.1 \\  
2 &2.2&0.55& 10.97 $\times$ 7.48 & 70.1 \\
2 &5.0& 0.055 & 6.11 $\times$ 4.93 &72.1  \\
2 &8.4& 0.11 & 3.42 $\times$ 2.77 &70.0   \\
\tableline
\end{tabular}
%% Any table notes must follow the \end{tabular} command.
\tablenotetext{a}{The rms noise levels near the reference position used in the JIVE VLBI correlation.}
%generated with the \LaTeX\ table environment}
%\tablenotetext{b}{Yet another sample footnote for table~\ref{tbl-2}}
%\tablenotetext{c}{Another sample footnote for table~\ref{tbl-2}}
%material to a table.}
%\end{center}
\end{table}
%------------------------ TABLE 3--------------
\begin{table}
\small
%\begin{center}
\caption{Positions of VLBI Components in NGC 6240 \label{tbl3}}
\begin{tabular}{lcccc}
\tableline\tableline
  Component& \multicolumn{2}{c}{Coordinate} & \multicolumn{2}{c}{Accuracy}\\
\cline{2-3}\cline{4-5} 
        &$\alpha$(J2000)   & $\delta$(J2000) &$\Delta\alpha$cos$\delta$$^{\dag}$ & $\Delta\delta$$^{\dag}$ \\
Frequency& 16$^{\rm h}$52$^{\rm m}$ & 02$\degr$24$\arcmin$&(mas)&(mas) \\ 
\tableline
Northern nucleus (N2) &$^{\rm s}$&$\arcsec$& \\
\tableline
  1.6 GHz  &58.9242 &04.779 &0.9 & 0.7  \\
  2.2 GHz&58.9240&04.777& 1.2&0.9\\
  5.0 GHz (Oct 2003) & 58.9240&04.776&0.6 &0.6\\
  5.0 GHz (Jun 2009) & 58.9240&04.776 & 0.6&0.6\\
  8.4 GHz  & 58.9240& 04.776 &0.6 &0.6\\
\tableline
Southern nucleus (N1)&&& \\
\tableline
1.6 GHz  &58.8903 &03.351& 1.1 &1.1  \\
5.0 GHz (Oct 2003) &58.8901 & 03.348 &0.7 &0.6\\
  5.0 GHz (Jun 2009) &58.8902 & 03.350 & 0.6&0.6\\
\tableline
Radio source 1 (RS1) & &  & \\   
\tableline
 5.0 GHz (Oct 2003) &58.8886 &03.326 &0.7&0.6\\
 5.0 GHz (Jun 2009) &58.8885 & 03.326 &0.6&0.6\\
8.4 GHz  &58.8885 &03.327  &0.6 &0.6\\
\tableline
Radio source 2 (RS2) & &  & \\   
\tableline
1.6 GHz &58.8991  &03.607   &2.0 &1.3\\
5.0 GHz (Oct 2003) &58.8993  & 03.602  &0.8 & 1.3\\
5.0 GHz (Jun 2009) &58.8994  & 03.602  &0.6 &0.6 \\
\tableline
\end{tabular}
%% Any table notes must follow the \end{tabular} command.
\tablenotetext {\dag}{The position error in right ascension ($\Delta\alpha$cos$\delta$) and declination ($\Delta\delta$).}
%generated with the \LaTeX\ table environment}
%\tablenotetext{b}{Yet another sample footnote for table~\ref{tbl-2}}
%\tablenotetext{c}{Another sample footnote for table~\ref{tbl-2}}
%\tablecomments{Units of R.A. (right ascension) are seconds, and units of Dec. (declination)
%are degrees, acrminutes, and arcseconds.}
%\tablecomments{Units of R.A. (right ascension) are hours, minutes, and seconds, and units of Dec. (declination)
%are degrees, acrminutes, and arcseconds.}
%material to a table.}
%\end{center}
\end{table}
%
%%------------------------ TABLE 4--------------
\begin{center}
\begin{deluxetable}{cccccccccccc}
%\begin{center}
\tablecolumns{12}
\tabletypesize{\scriptsize}
\tablewidth{0pc}
\tablecaption{Component Flux Densities of NGC 6240 \label{tbl4}}
%\begin{tabular}
\tablehead{
& \multicolumn{3}{c}{Epoch 1 (EVN;Oct$-$Nov 2003)}& &\multicolumn{4}{c}{Epoch 2 (EVN;June 2009)}&&\multicolumn{1}{c}{MERLIN}\\
  \cline{2-4} \cline{6-9} \cline{11-11}\\
%Name&1.6 GHz &5.0 GHz&$\alpha_{1.6-5.0}$&&2.2 GHz&5.0 GHz&8.4 GHz&$\alpha_{5.0-8.4}$&&5.0 GHz&8.4 GHz\\
Name&$S_{\rm {1.6}}$ &$S_{\rm {5.0}}$&$\alpha^{\rm {1.6}}_{\rm {5.0}}$&&$S_{\rm{2.2}}$&$S_{\rm{5.0}}$&
$S_{\rm{8.4}}$&$\alpha^{\rm{5.0}}_{\rm{8.4}}$&&$S_{\rm{5.0}}$\\
  (1)&(2)&(3)&(4)&&(5)&(6)&(7)&(8)&&(9)
}
\startdata
N1&2.09&2.63&0.20&&$<$ 2.75&2.85&$<$0.55&$<$-3.2&&36.8\\
N2&3.95&5.99&0.37&&4.59&5.79&1.78&-2.3 &&16.1 \\
RS1& \nodata& 2.34& \nodata &&$<$2.75&3.00&3.57&0.34&&\nodata \\
RS2&1.07&0.38&-0.91&&$<$2.75&0.59&$< $0.55&$<$-0.14&&\nodata \\ 
\hline \\
&&&&&VLBA&&&&\\
\hline
& \multicolumn{2}{c}{Epoch 1 (Aug 1999)}& &\multicolumn{2}{c}{Epoch 2 (Apr$-$May 2000)}&&\multicolumn{2}{c}{Epoch 3 (Dec 2000$-$Jan 2001)}\\
\cline{2-3} \cline{5-6} \cline{8-9}\\
&$S_{\rm{1.7}}$ &$S_{\rm{2.4}}$ && $S_{\rm{1.7}}$& $S_{\rm{2.4}}$&& $S_{\rm{1.7}}$&$S_{\rm{2.4}}$\\
 (10) & (11)& (12)    &&(13)   &(14) &&(15) &(16)\\
S1& \nodata& 0.63$\pm$0.13& &0.62$\pm$0.06&0.59$\pm$0.10&&0.40$\pm$0.08&0.47$\pm$0.08 \\
RS3&\nodata &\nodata & &\nodata&\nodata&&\nodata&0.46  \\
RS4&\nodata &\nodata & &0.44&0.29&\nodata&\nodata&\nodata \\
\enddata
\tablecomments{Flux densities are in mJy. Col. (1): Source names from Baan, Hagiwara, \& Hofner (2007) for N1 
(southern nucleus) and N2 (northern nucleus). RS1 and RS2
are newly detected sources in these EVN observations. Flux density errors measured 
by the EVN are estimated at $\sim$ 10\%.
Col. (2),(3): The 1.6 GHz and 5.0 GHz flux densities at Epoch 1 as measured by the EVN.
Col. (5)-(7): The 2.2 GHz, 5.0 GHz, and 8.4 GHz flux densities at Epoch 2 as measured by 
the EVN.
Col. (4),(8): The spectral indices between 1.6 GHz and 5.0 GHz (Epoch 1) and between 5.0 GHz 
and 8.4 GHz (Epoch 2).
Col. (9): Data taken from MERLIN observations at 5.0 GHz in \citet{bes01}. All flux density upper limits 
are 5 $\sigma$.
Col. (10): Data taken from VLBA observations in \citet{gall04}. RS3 ($\approx$0$\arcsec$.02 northwest 
of the southern nucleus N1) and RS4 ($\approx$0$\arcsec$.03 northwest of the 
northern nucleus N2) are newly defined in this article.
Col. (11)-(16): The 1.7 GHz and 2.4 GHz flux densities as measured by the VLBA \citep{gall04}.
}
\end{deluxetable}
\end{center}
\begin{center}
\begin{deluxetable}{cccccccccccccccc}
%\tablecolumns{16}
\tabletypesize{\scriptsize}
\tablewidth{0pc}
\rotate
\tablecaption{Source sizes of N1, N2, RS1, and RS2 \label{tbl5}}
\tablehead{
&\multicolumn{3}{c}{N1} && \multicolumn{3}{c}{N2} && \multicolumn{3}{c}{RS1} & & \multicolumn{3}{c}{RS2}  \\
\cline{2-4} \cline{6-8}  \cline{10-12} \cline{14-16}\\
(1)&(2)&(3)&(4)&&(5)&(6)&(7)&&(8)&(9)&(10)&&(11)&(12)&(13) \\
Frequency (Epoch) &(mas$^2$)&(pc$^2$)&($\degr$)&&(mas$^2$)&(pc$^2$)&($\degr$)&&(mas$^2$)&(pc$^2$)&($\degr$)&&(mas$^2$)&(pc$^2$)&($\degr$) \\
% & &&&&&&&&&&&&&& \\
}
\startdata
1.6 GHz (Oct 2003) & 17.6$\times$12.6&8.4$\times$6.0 & 43&&39.3$\times$13.5 &18.7$\times$6.4 & 5 &&$-$&$-$&$-$&&15.8$\times$7.0&7.5$\times$3.3 &169   \\
5.0 GHz (Oct 2003)  &9.2$\times$5.9&4.4$\times$2.8&137&&19.5$\times$6.6 & 9.3$\times$3.2&98 &&9.2$\times$3.9&4.4$\times$1.9& 80 &  &9.2$\times$6.1&4.4$\times$2.9&140       \\
5.0 GHz (Jun 2009)  & 7.6$\times$5.2& 3.6$\times$2.5&161 && 12.4$\times$3.4&5.9$\times$1.6 &101  &&3.6$\times$1.9 &1.7$\times$0.9&70 &  &2.5$\times$2.9&1.2$\times$1.4&128   \\
8.4 GHz (Jun 2009) & $-$&$-$ &$-$ && 6.6$\times$1.4& 3.1$\times$0.7 & 85 &&3.2$\times$1.3&1.5$\times$0.6&58  &  &$-$&$-$&$-$     \\
\enddata
\tablecomments{Source sizes are upper limit values and they are deconvolved using the VLBI synthesized CLEAN beam. 
Col. (1): The observing frequency and Epoch. Col. (2),(5),(8),(11): Major axis $\times$ minor axis values, in milli-arcsecond (mas). 
Col. (3),(6),(9),(12):  Major axis $\times$ minor axis values, in parsecs (pc).
Col. (4),(7),(10),(13): The position angles of the sources.
}
\end{deluxetable}
\end{center}
%------------------------ TABLE 6-------------
%\begin{center}
\begin{table}
%\small
%\begin{center}
\caption{Properties of the VLBI sources N1, N2, RS1, and RS2 \label{tbl6}}
\begin{tabular}{lcccc}
\tableline\tableline
Source    &T$_{\rm b,5GHz}$ &P (5.0 GHz)&T$_{\rm b,8.4GHz}$ & P (8.4 GHz) \\
&($\times$ 10$^{6}$ K)&($\times$ 10$^{21}$ W Hz$^{-1}$)&($\times$ 10$^{6}$ K)& 
($\times$ 10$^{21}$ W Hz$^{-1}$) \\
\hline
Northern nucleus (N2)&10.0& 2.9 & 4.9 & 1.7 \\
Southern nucleus (N1)&5.2   &  1.7 & $<$1.5  & $<$0.70$^{\dag}$  \\
Radio source 1 (RS1)& 31.6 & 3.9  &21.8 & 4.1\\
Radio source 2 (RS2) &5.9  & 0.75 & $<$1.5 &$<$0.70$^{\dag}$  \\
\hline
\end{tabular}
\tablecomments{These values are calculated from data obtained in June 2009. {\dag}{5 $\sigma$ upper limit}}
%% Any table notes must follow the \end{tabular} command.
%\tablenotetext{a}{Deconvolved by the VLBI synthesized CLEAN beam}
%generated with the \LaTeX\ table environment}
%\tablenotetext{b}{Yet another sample footnote for table~\ref{tbl-2}}
%\tablenotetext{c}{Another sample footnote for table~\ref{tbl-2}}
%material to a table.}
%\end{center}
\end{table}
%\end{center}
%
%
%------------------------ TABLE 7--------------
\begin{center}
\begin{deluxetable}{lcc}
%\begin{center}
\tablecolumns{3}
\tabletypesize{\scriptsize}
\tablewidth{0pc}
\tablecaption{Modeling fit parameters for the nuclei N1 and N2  \label{tbl7}}
%\begin{tabular}{lclc|c}
%\tableline\tableline
\tablehead{
Parameter~~~~~~~~~~~&\multicolumn{2}{c}{Value}
}
\startdata
%&\multicolumn{2}{c}{Value} \\
%\cline{2-3} 
Free-free absorption model&N1&N2 \\
\tableline 
$S_{0}$& 25.4 $\pm$ 35.7  mJy  & 30.4 $\pm$ 60.8  mJy\\
$\alpha$&  -1.4 $\pm$ 0.8   & -1.1 $\pm$ -1.1 \\
$\tau_{\rm {ff}}$& 5.3 $\pm$ 3.5    &4.5 $\pm$ 4.9 \\
Reduced $\chi^{2}$ & 0.59  & 3.8 \\
\tableline\\
Synchrotron self-absorption model& N1&N2\\
\tableline
$S_{0}$&  0.39$\pm$ 0.07  mJy   &  0.74$\pm$ 0.22  mJy\\
$\alpha_{0}$&  -3.2 $\pm$ 2.3     &-2.40 $\pm$ 1.97 \\
$\tau_{\rm {s}}$&  1349 $\pm$ 5110    &400 $\pm$ 1326 \\
Reduced $\chi^{2}$ & 0.43   & 2.9 \\
\enddata
%\end{tabular}
%\tablenotetext{b}{Yet another sample footnote for table~\ref{tbl-2}}
%\tablenotetext{c}{Another sample footnote for table~\ref{tbl-2}}
%material to a table.}
%\end{center}
\end{deluxetable}
\end{center}
%
%
%\clearpage
%
%------------------------ FIGURE 1--------------
% 22GHz Overview VLA map

%------------------------ FIGURE 2--------------
%%%%%%%%%%%%%%%%%%%%%%%%%%%%%%%%%%%%%%%%%%%%%%%
% N1: SOUTHERN NUCLEUS AT 1.6, 5.0, and 8.4 GHz    IN  EP1 and EP2
%%%%%%%%%%%%%%%%%%%%%%%%%%%%%%%%%%%%%%%%%%%%%%%
%
%------------------------ FIGURE 3--------------
%%%%%%%%%%%%%%%%%%%%%%%%%%%%%%%%%%%%%%%%%%%%%%%%%%%%%%
% N2: NORTHERN NUCLEUS AT 2.2, 5, and 8.4 GHz   IN  EP1 and EP2
%%%%%%%%%%%%%%%%%%%%%%%%%%%%%%%%%%%%%%%%%%%%%%%%%%%%%%%%%
%------------------------ FIGURE 4--------------
%%%%%%%%%%%%%%%%%%%%%%%%%%%%%%%%%%%%%%%%%%%%%%%%%
% RS2: Northeast of N1
%%%%%%%%%%%%%%%%%%%%%%%%%%%%%%%%%%%%%%%%%%%%%%%%%
%
%------------------------ FIGURE 5-------------
%%%%%%%%%%%%%%%%%%%
% SED of N1 and N2
%%%%%%%%%%%%%%%%%%%

%------------------------ FIGURE 6--------------
% Right curve of RS1

\end{document}